\documentclass[11pt]{article}
\usepackage{hyperref, graphicx, url, pslatex, paralist}
\usepackage[margin={2.7cm,2.7cm}]{geometry}

\begin{document}

\title{\vspace{-2cm}An Academic's Observations from a Sabbatical at Google.}
\author{Adam Barker, University of St Andrews, UK \\ \texttt{adam.barker@st-andrews.ac.uk}}
\date{}
\maketitle
\begin{abstract} \centering \emph{How experiences gained in industry can improve academic research and teaching.}
\end{abstract}

I have spent the majority of my computing career in an academic environment, with a Ph.D. in Informatics from the University of Edinburgh (2007), postdoctoral experience at the Universities of Edinburgh, Oxford, and Melbourne (2007--2010), followed by a faculty position at the University of St Andrews in Scotland. However, several questions had always played on my mind: ``What is industry really like?", ``Am I working on the right problems?", ``How can I get people to use my research?". To answer some of these questions I applied to the Google Visiting Faculty program~\cite{visitng} where I spent nine months as a Visiting Scientist in Mountain View, CA, as part of the Borgmaster team. Borg~\cite{borg} is Google's cluster management framework, which runs hundreds of thousands of jobs, across a number of clusters each with up to tens of thousands of machines. This Viewpoint summarizes some of my experiences working in software engineering at Google, with each subsection derived from observations inspiring the core lessons I am bringing back to academia in order to improve both research and teaching.

\medskip
\textbf{Production software engineering.} One of the most useful experiences during my time at Google was writing production-quality code. This is not something I had a great deal of experience with, coming primarily from an academic background, where the output is usually research or prototype code. The differences and motivation for code between academia and industry are stark: in an academic environment, code is primarily written for the purposes of demonstrating a proof of concept and is often of low quality and even disposed of after a paper deadline or project finishes. The reason behind this is not that academics cannot write code, but that the code is often a means to an end. The new knowledge and insights, written up as a research paper, are the artifacts that researchers value. In industry, code is written for a group of end users and developers who must maintain and extend its functionality. It must, therefore, pass rigorous testing and code review before it is checked in and made available to the rest of the company, and eventually the outside world.

Although I had previously taught software engineering, as an academic I found the process of designing, testing, and having code reviewed to be priceless: it highlighted how the software engineering process works in reality at a large organization. Google's tooling is primarily in-house and involves trunk-based development using Piper for version control, Critique for code review, Tricorder for static analysis, and ``presubmit" infrastructure that provides automated testing and analysis of changes before they are added to the code base~\cite{google}. The learning curve was steep at first, especially because the focus of my academic job is not on software engineering, but instead on writing, research management and supervision, and teaching.

I highly recommend that all academics working in practical or applied areas gain an understanding of software engineering processes first-hand. The benefits to the university may appear subtle at first but could lead to more effective and realistic university-level software engineering courses and therefore increased potential employability of graduates. Furthermore, if you apply these engineering principles (production code, code reviews, bug tracking, design documents, and so forth) to the appropriate parts of your own research lab's work then ultimately this could lead to higher-quality code, which in turn will allow more researchers to utilize and build upon your work.

\medskip
\textbf{Making an impact in industry.} My time in industry highlighted a set of interesting insights about how to make academic research more relevant to industry. A typical computer system's research paper consists of a research problem underlined by a real-world use case; the research contributions conveyed through a new idea or concept; an implementation and evaluation based on either a prototype or simulation; and a set of assumptions and constraints about the environment the proposed system operates in, as well as the prior knowledge it may have.

After spending time at Google, I can conclude that in general most academic implementations will never get picked up by industry because the artifact itself will not fit in with the complex ecosystems that exist in most large companies. There are obvious exceptions to this, for example, Spark and Mesos, which started as academic research projects with a lot of input from industry. New concepts and ideas can quite easily get picked up by industry, but quite often the reason they do not translate is that the assumptions the research makes are incompatible with the real world. As an example, scheduling research that makes unrealistic assumptions about the environment the scheduler must operate in, covering factors such as prior knowledge, resource availability, and time available to make a scheduling decision. Or academic cloud computing research~\cite{hotcloud} based on simulation, or experiments on small toy private cloud deployments, which are non-representative of real-world data centers.

By taking time out of academia it is likely that academics can improve their knowledge and understanding of the real-world problems faced by industry, and the complex production environments and constraints systems must operate in. If research properly addresses these two criteria it is more likely to gain traction in an industrial setting. It is worth acknowledging that a lot of academic computer science research is more basic and long-term, and should not be driven by today's industry issues.

\medskip
\textbf{Working on the ``right" problems.} Academia's primary role is to advance knowledge, shed light on complex problems and to gain understanding, not just to engineer solutions~\cite{sns}. The difficulty within academia is to determine which problems are the ``right" problems to work on, so the work has an impact. A common problem is that academic research can become disjoint from industry advances within the field, and we end up with complex solutions that are looking for real-world problems. This is one of the motivations for Google's hybrid research model~\cite{hybrid} where the majority of research is embedded directly into an engineering team.

The best way to work out which problems academics should be working on as a community is to go out and actually talk to industry about the real problems they face. To do this we need simpler, cleaner collaboration mechanisms so that industry and academia can work together more effectively. There are several schemes I have encountered that facilitate these collaborations, examples include the Google Visiting Faculty scheme and Royal Society Industry Fellowships~\cite{royal}, but as a community, we need more. There are university research labs that do this really well, but sadly they are few and far between. A good example is the Algorithms Machines and People (AMP)~\cite{amp} at UC Berkeley, where I had the good fortune to spend a summer. The other option is to send your Ph.D. students to intern in industry: they will then return with fresh ideas and new perspectives. Annual or bi-annual industry retreats are another great way of receiving industry input into university-led research projects.

\medskip
\textbf{Running a team.} As an academic, I am used to having my own private office. The opposite was true at Google, where everyone sat in large open spaces with lots of breakout rooms for meetings and calls. At first, I found this distracting but quickly adjusted to the benefits of this collaborative, team-focused space offered. Many of the processes in Google were also streamlined: individual meetings with my line manager and mentor were typically half an hour, with a weekly all-hands meeting that lasted up to an hour, email was minimized, and documents collaborative.

After experiencing this different way of working I am attempting to run my research group much more like a software team~\cite{score}. I have moved all of my Ph.D. students into a single shared systems research lab. I typically run a stand-up meeting once a week with students and postdocs -- in these meetings everyone discusses progress from the previous week and sets clear objectives for the week ahead. Longer, more detailed meetings to discuss research are then organized on demand, but I try to limit them to half an hour. We finalize, or ``ship" papers and associated code by specified deadlines, usually driven by an upcoming conference. This allows us to iterate and provide feedback on research ideas quickly, and avoid going down too many research dead ends. We use online collaborative tools such as Trello for task management, and Slack for communication to reduce email overhead. Papers and code, which can be reviewed by other members of the lab, are pushed into shared repositories available to the whole research group.

In summary, the experience of stepping outside a purely academic environment has opened my eyes to a different way of conducting computer science research. I would highly recommend this sort of experience to academics. You might not get a paper out of it, but by taking some time out of the daily research grind and getting your hands dirty in software engineering you will walk away with a better understanding of how industry engineering and management processes really work. These experiences can inevitably feed into and improve academic research and teaching. My advice to academics wanting to take a sabbatical in industry would be to find a particular group inside a company you want to work with and propose a collaborative research project that could potentially improve their systems, processes, or business. Dust off your coding skills as they may be a little rusty, and practice designing, writing, testing, and committing code. To industry, I would recommend finding simple, clean ways of working with academia on the difficult engineering challenges you face.

\medskip
\textbf{Acknowledgements.} Thank you to John Wilkes and Steven Hand for their contributions to this Viewpoint, and more broadly for their time and input during my Visiting Faculty residency. In addition, thank you to the entire Borgmaster and Visiting Faculty teams at Google.

\bibliographystyle{abbrv} \bibliography{cacm}

\end{document}